\documentclass{agujournal}
\usepackage{hyperref}

\journalname{Geophysical Research Letters}

\begin{document}

\title{Saturn's northern aurorae at solstice from HST observations coordinated with Cassini's Grand Finale}

\authors{L.~Lamy\affil{1}, R.~Prang\'e\affil{1}, C.~Tao\affil{2}, T.~Kim\affil{3}, S.~V.~Badman\affil{4}, P.~Zarka\affil{1}, B.~Cecconi\affil{1}, W.~S.~Kurth\affil{5}, W.~Pryor\affil{6}, E.~Bunce\affil{7}, A.~Radioti\affil{8}}

\affiliation{1}{LESIA, Obs. de Paris, PSL, CNRS, UPMC, Univ. Paris Diderot, Meudon, France.}
\affiliation{2}{National Institute of Information and Communications Technology, Tokyo, Japan.}
\affiliation{3}{Center for Space Plasma and Aeronomic Research, University of Alabama, Huntsville, USA.}
\affiliation{4}{Department of Physics, Lancaster University, Lancaster, UK.}
\affiliation{5}{Department of Physics and Astronomy, University of Iowa, Iowa City, USA.}
\affiliation{6}{Department of Science, Central Arizona College, Coolidge, USA.}
\affiliation{7}{Department of Physics and Astronomy, University of Leicester, Leicester, UK.}
\affiliation{8}{Space Science, Technologies and Astrophysics Research Institute, Li\`ege, Belgium.}

\correspondingauthor{L. Lamy}{laurent.lamy@obspm.fr}


\begin{keypoints}
\item Saturn's northern UV aurorae at solstice were sampled from HST observations coordinated with Cassini's Grand Finale.
\item The observed aurorae are highly variable with powerful events, radiating up to 120~GW, controlled by solar wind and planetary rotation.
\item The average auroral brightness strongly varies with LT with two maxima at dawn (previously known) and pre-midnight (newly identified).
\end{keypoints}

\begin{abstract}

Throughout 2017, the Hubble Space Telescope (HST) observed the northern far-ultraviolet aurorae of Saturn at northern solstice, during the Cassini Grand Finale. These conditions provided a complete viewing of the northern auroral region from Earth and a maximal solar illumination, expected to maximize the ionosphere-magnetosphere coupling. In this study, we analyze 24 HST images concurrently with Cassini measurements of Saturn's Kilometric Radiation and solar wind parameters predicted by two MHD models. The aurorae reveal highly variable components, down to timescales of minutes, radiating 7 to 124$\pm11$~GW. They include a nightside-shifted main oval, unexpectedly frequent and bright cusp emissions and a dayside low latitude oval. On average, these emissions display a strong Local Time dependence with two maxima at dawn and pre-midnight, the latter being newly observed and attributed to nightside injections possibly associated with solstice conditions. These results provide a reference frame to analyze Cassini {\it in situ} measurements, whether simultaneous or not.

\end{abstract}

\section{Introduction}
Saturn's aurorae have been intensively observed from Earth with the Hubble Space Telescope (HST) in the far-ultraviolet (FUV) mainly using the Space Telescope Imaging Spectrograph (STIS) and the Advanced Camera for Surveys (ACS). Many of these observations were coordinated with {\it in situ} measurements from the Cassini spacecraft, including its Ultraviolet Imaging Spectrometer (UVIS), during its orbital tour from 2004 to 2017. Recent reviews summarize our current understanding of kronian auroral processes \citep[and refs therein]{Kurth_09,Badman_SSR_15,Stallard_14}.

The UV aurorae are the neutral atmospheric response of the prominent H and H$_2$ species to precipitations of electrons energized in the magnetosphere. The energy of primary electrons, measured by various methods based on spectroscopic HST/STIS and Cassini/UVIS measurements ranges from a few keV to a few tens of keV \citep{Gustin_Icarus_17}. The kronian aurorae decompose into a variety of components, tentatively listed by \citet{Grodent_SSR_15}, driven by different acceleration processes and underlying current systems. We hereafter restrict ourselves to four broad categories.

The dominant auroral emission is a circumpolar main oval, whose intensity and location significantly vary with time. It was early found to be associated with Saturn's Kilometric Radiation (SKR) \citep{Kurth_Nature_05} and with strong upward field-aligned currents located slightly equatorward of the open-closed field line boundary \citep{Bunce_JGR_08,Belenkaya_AG_08,Hunt_JGR_15b}. It typically radiates a few tens of kilo-Rayleighs (kR, local photon flux per pixel) and a few tens of GW (total power radiated by the whole auroral region), although differing definitions of these quantities used in the literature prevent us from cross-comparing them. The quiet main oval is a quasi-circular narrow faint ring of emission near $72-75^\circ$ northern latitudes. By contrast, magnetospheric compressions driven by interplanetary shocks trigger bright auroral storms typically lasting for $\sim$1.5~planetary rotations ($1.5\times\sim$10.7~h), with a significant part of the main oval expanding toward high latitudes \citep{Prange_Nature_04,Clarke_Nature_05,Clarke_JGR_09,Meredith_JGR_14b}. Longitudinally extended intensifications along the undisturbed oval phased with SKR were alternately related to rotationally-modulated nightside injections \citep{Jackman_JGR_09,Mitchell_PSS_09,Nichols_GRL_10a,Lamy_JGR_13}. The main oval additionally hosts a variety of smaller-scale transient and/or sub-corotating hot spots \citep{Radioti_JGR_09,Grodent_JGR_11,Meredith_JGR_13,Radioti_JGR_15}. On average, its brightness strongly varies with Local Time (LT) with a main maximum at dawn \citep{Lamy_JGR_09,Carbary_JGR_12}, a peculiarity of Saturn's aurorae.

Cusp aurorae have also been occasionally identified as emissions radiating a few GW and up to 50~kR, varying on timescales of hours, and confined close to noon either along the main oval or poleward of it depending on the orientation of the interplanetary magnetic field \citep{Gerard_JGR_05,Meredith_JGR_14a,Palmaerts_JGR_16,Kinrade_JGR_17}. Such signatures are sometimes associated with duskside bifurcations of the main emission, similarly attributed to dynamical dayside reconnection \citep{Radioti_JGR_11}. A faint secondary oval, $\sim2$~kR bright, has also been identified on the southern nightside in HST/STIS and then Cassini/UVIS images equatorward of the main one \citep{Grodent_JGR_10,Lamy_JGR_13,Radioti_JGR_17}. It appeared as a few-degrees-wide ring near $-67^\circ$ southern latitude, which \citet{Grodent_JGR_10} attributed to the precipitation of suprathermal electrons from the middle magnetosphere rather than to a field-aligned current system. Finally, a last important auroral feature consists of a spot at the magnetic footprint of Enceladus driven by the planet-satellite interaction. It was identified as a 1~kR bright emission near $+64.5^\circ$ latitude in only three Cassini/UVIS images so far \citep{Pryor_Nature_11}. 

In the frame of Cassini's Grand Finale, HST/STIS regularly observed Saturn's northern aurorae throughout 2017, during northern summer solstice (reached on 24 may). These conditions offered the best achievable HST viewing of the northern auroral region. They also provided maximal solar illumination, i.e. maximal northern ionospheric conductivity and thus maximized ionosphere-magnetosphere coupling through the current systems driving most of the kronian aurorae. The HST observations were carefully coordinated with {\it in situ} measurements of the Cassini spacecraft within the auroral region. In this article, we analyze 24 HST/STIS images concurrently with Cassini SKR observations and propagated solar wind (SW) parameters. The dataset is presented and analyzed in sections \ref{data}-\ref{results}. Results are then discussed in section \ref{discussion}. Details on the HST data processing and supplementary Figures are provided in the supplementary material. 

\section{Dataset}
\label{data}



\subsection{HST/STIS observations}
\label{hst}

During the Cassini's Grand Finale, STIS observed Saturn's FUV auroral emissions during 25 HST orbits distributed throughout 2017. These were scheduled when Cassini was planned to traverse SKR sources, themselves colocated with layers of auroral field-aligned upward currents \citep{Lamy_Science_18}. Each orbit included one $\sim44$~min long time-tagged exposure. The time-tag mode provides the arrival time of photons recorded on the STIS MAMA (Multi-Anode Microchannel Array) detector at a 125~microsec resolution and thus enables us to track dynamics at timescales shorter than the exposure (see supplementary Figure S1 and Animation S1). Out of 25 orbits, 24 acquired 1024$\times$1024~pix images at 0.00247 arcsec.pix$^{-1}$ resolution with the Strontium Fluoride filter F25SrF$_2$ (148~nm central wavelength, 28~nm FWHM) which rejects wavelengths shortward of 128~nm and notably the H Ly-$\alpha$ line. One orbit was also used to slew the northern auroral region with the 0.5~arcsec slit and the G140L grating. In this study, we focus on the analysis of the images, processed and translated into brightnesses and power radiated over the 70-180~nm H$_2$ bands (see supplementary material).

\subsection{Cassini/RPWS data and solar wind models}
\label{other}

Cassini quasi-continuous observations of the Radio and Plasma Wave Science experiment (RPWS) \citep{Gurnett_SSR_04} were used to monitor the activity of Saturn's Kilometric Radiation between a few kHz and $\sim$1100~kHz. We derived flux densities normalized to 1~AU observing distance and power integrated over 10-1100~kHz for comparison purposes with previous studies \citep{Lamy_JGR_08a}. The normalization assumes a source-observer distance equal to the planet-observer one, an assumption which is less and less valid for closer Cassini-Saturn distances, but fair enough to assess typical intensities. We also used northern SKR phases, derived as described in \citep{Lamy_PRE7_11} from the most recent northern SKR period ($\sim10.8$~h throughout 2017) \citep{Lamy_PRE8_17}.

SW parameters were numerically propagated from the Earth's orbit out to Saturn with two magneto-hydrodynamic (MHD) codes. The Tao 1D model was developed for the jovian case \citep{Tao_JGR_05} and extrapolated to Saturn's orbit (e.g. \citep{Kimura_JGR_13}). The Multi-scale Fluid-kinetic Simulation Suite (MS-FLUKSS) \citep{Pogorelov_14} is a 3D model validated in the outer heliosphere thanks to {\it in situ} plasma measurements of Ulysses, Voyager and New Horizons \citep{Kim_ApJ_16}. The input parameters are, for both models, near-Earth SW {\it in situ} observations provided by either NASA/GSFC's OMNI 1h averaged data obtained from Wind measurements \citep{King_JGR_05} and/or Stereo-A measurements instead. The uncertainty depends on the derived parameters and on the angular separation between the Earth and Saturn and gradually increases from opposition. For angular separations less than 90$^\circ$, which provided a fair coverage of 2017 through complementary Omni and Stereo-A inputs, the typical uncertainty on the timing of dynamic pressure fronts is estimated to be less than $\pm35$~h, according to previous results from another 1D model \citep{Zieger_JGR_08}, whose results are not available for this study.


\section{Results}
\label{results}

\subsection{A variety of variable components}
\label{components}

\begin{figure*}[ht!]
\centering
\noindent\includegraphics[width=35pc,angle=-0]{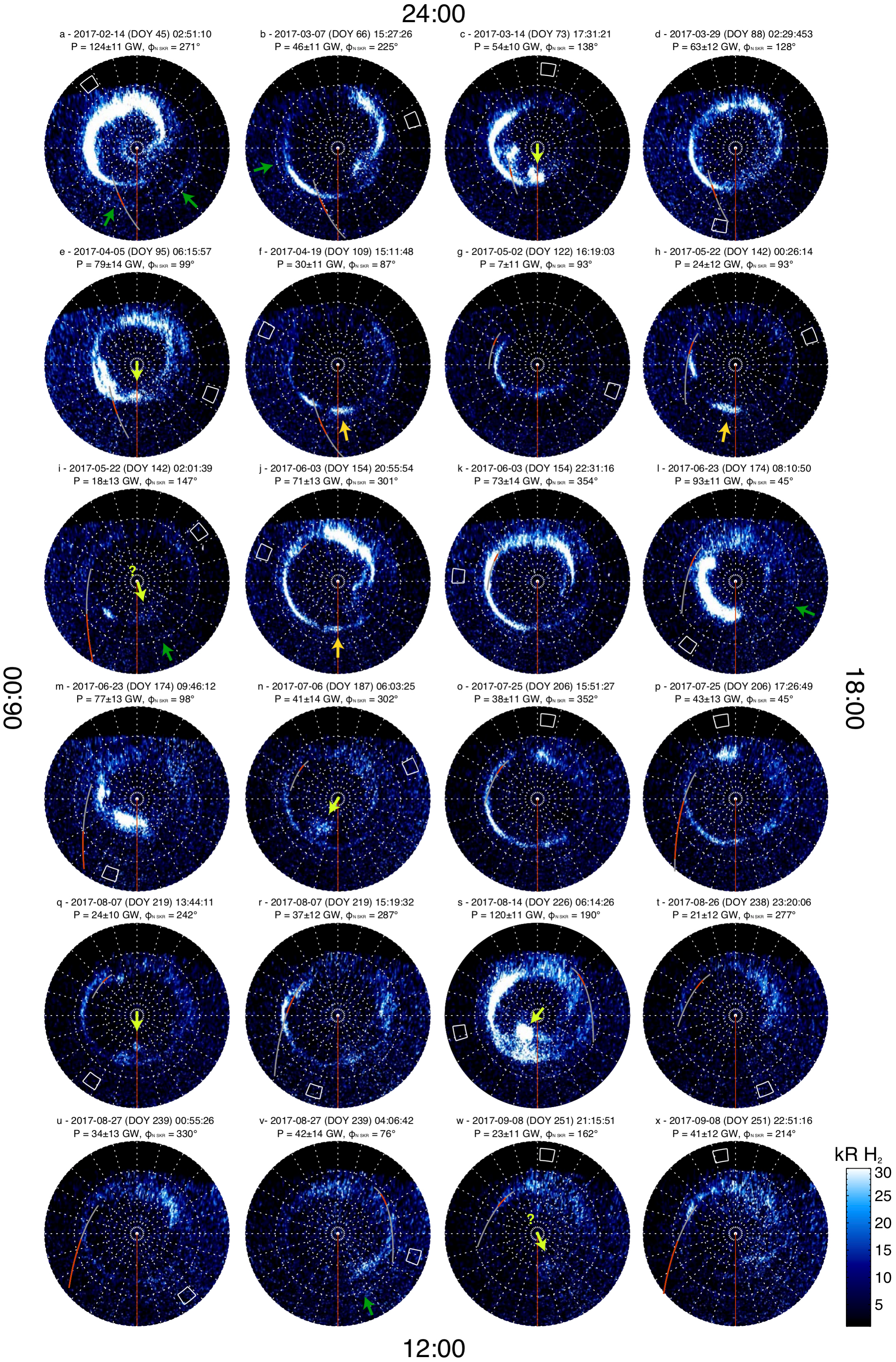}
\caption{Polar projections of HST/STIS images of northern kronian aurorae in 2017 versus LT (red meridian = noon), projected at 1100~km altitude, with light-travel corrected observing time, northern SKR phase and total power radiated in the H$_2$ bands above each image. Yellow (orange) arrows indicate plausible cusp emissions poleward of (along) the main oval. Green arrows indicate low-latitude emission. White boxes indicate the Enceladus magnetic footprint. The red (gray) curves plot the Cassini magnetic footprint during ($\pm2$~h aside) each HST exposure.}
\label{fig1}
\end{figure*}

Figure \ref{fig1} displays polar projections of all STIS images, labelled a to x, versus LT. Images h-i, j-k, l-m, o-p, q-r, t-v and w-x were acquired along successive HST orbits. The observations reveal diverse aurorae with highly variable, down to timescales of minutes (see Figure S1 and Animation S1), localized features dominated in intensity by the main oval. The latter is an inhomogeneous circumpolar ring generally more intense at dawn and pre-midnight. The quiet oval is quasi-circular at 72-73$^\circ$ latitude (images n, u-v) with brightnesses reaching a few tens of kR. Whenever active, it reaches higher latitudes (up to $87-88^\circ$ in image a), often with a left-handed spiral shape (the oval develops counterclockwise from the pole). The brightest emissions exceed 150~kR (images a, l, s). Overall, half of the STIS images were acquired when the Cassini magnetic footprint simultaneously intercepted aurora (red curves).

Isolated features regularly observed near noon poleward of the main oval (yellow arrows) or along it (orange arrows) are then interpreted as cusp aurorae. In the latter case, this interpretation is not unambiguous as hot spots sub-corotating along the main oval can move through noon from dawn to dusk \citep{Meredith_JGR_13}. These events are often associated with duskside bifurcations of the main oval toward high latitudes (images b, e, j-k, n, s, x) accounting for the general left-handed spiral shape of the main emission. The association between bifurcations, suggesting dynamical lobe reconnection, and noon spots strengthens their interpretation as cusp emissions. Image s shows the brightest example around $10$:$30$~LT at $84^\circ$ latitude, persisting over 44~min and variable at timescales of minutes (see Animation S1 and Figure S1) with unusual brightnesses $\ge100$~kR, the largest ever reported.

A faint dayside oval equatorward of the main one appears in several images (a, b, l, v, green arrows) within $65-72^\circ$ latitude, sometimes in restricted LT sectors. While this secondary emission is distinct from the main one, we cannot exclude that it is reminiscent of ancient long-lived structures of the main oval which moved toward lower latitudes. Half of these examples correspond to active events when the main oval shifted to high latitudes. In image a, this secondary oval appears at $69-72^\circ$ from $06$:$00$~LT (or $00$:$00$~LT $\le70^\circ$) to $18$:$00$~LT, with peak brightnesses $\sim10$~kR. Interestingly, during this exposure, Cassini intercepted the secondary oval near $11$:$00$~LT during which MAG measurements \citep{Dougherty_SSR_04} of the azimuthal magnetic component (supplementary Figure S2) reveal successive small-scale abrupt gradients consistent with field-aligned current signatures, preceded $\sim$1~h earlier by a large positive gradient indicating a strong upward current layer associated with the poleward main emission (e.g. \citep{Bunce_JGR_08,Talboys_JGR_09,Talboys_JGR_11,Hunt_JGR_14,Lamy_Science_18}).

Finally, we could not detect any Enceladus footprint (white boxes).

\subsection{Transient enhancements and associated drivers}

\begin{figure*}
\centering
\noindent\includegraphics[width=35pc,angle=-0]{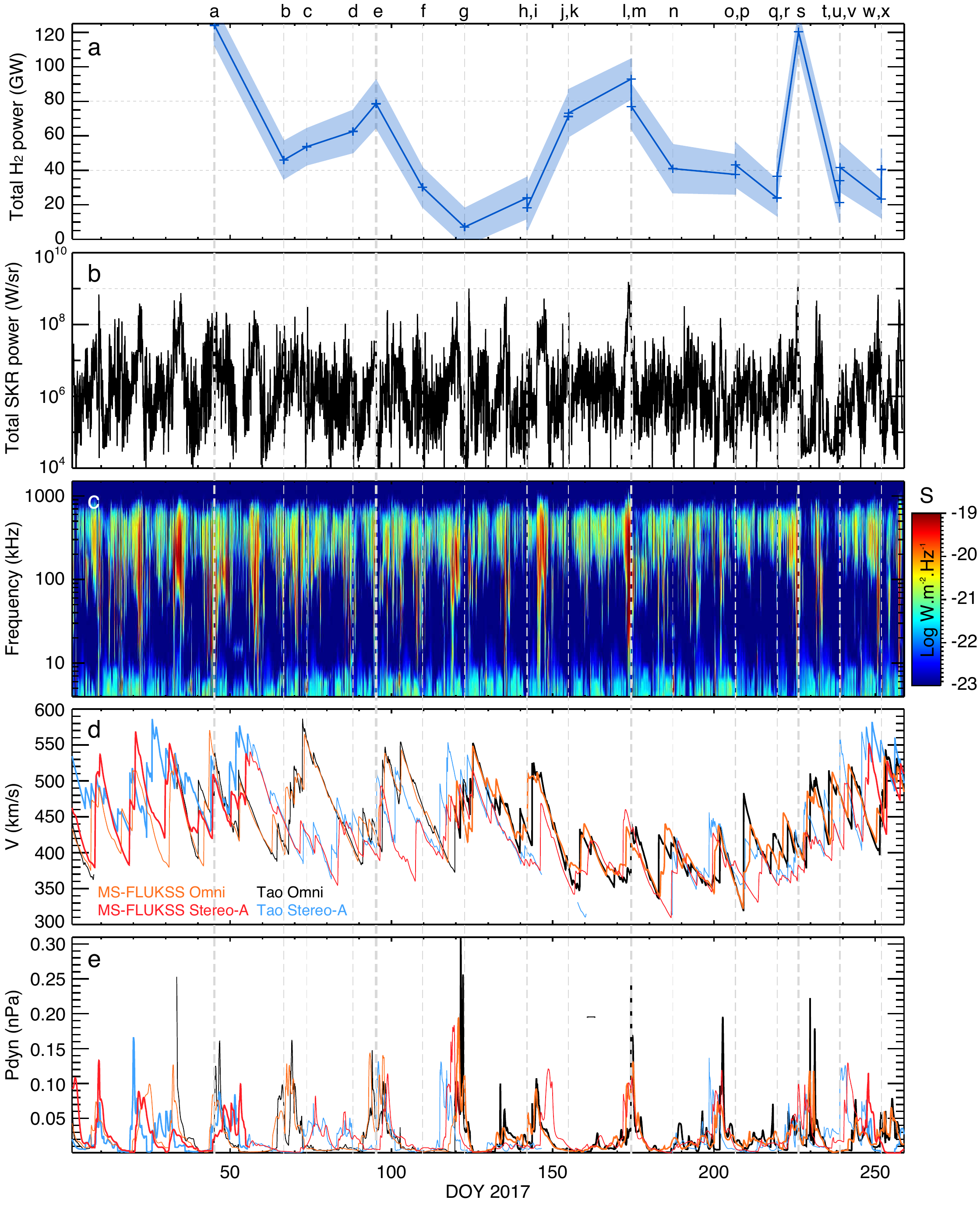}
\caption{(a) Total auroral power radiated in the H$_2$ bands. (b) Total SKR power over 10-1100~kHz at 90~s resolution derived from (c) Cassini/RPWS dynamic spectrum of spectral flux density. (d-e) Velocity and dynamic pressure propagated at Saturn by two MHD models, using either Omni or Stereo-A data inputs. Boldface portions indicated angular separation between Wind/Stereo-A and Saturn $\le90^\circ$. The double black arrow plots a $\pm35$~h error bar. }
\label{fig2}
\end{figure*}

Figure \ref{fig2}a plots the total auroral power radiated by H$_2$ as a function of time, the individual power values being listed in Figure \ref{fig1}. Overall, the auroral power strongly varies with time, with a factor of $\sim$10 between extremal values, weeks apart (s-t), and with a factor of $\sim$2 between consecutive orbits (t-v or w-x). Seven events showing active aurorae radiated power $\ge65$~GW, up to 124, 93 and $120\pm11$~GW for images a, l and s, resp. The latter includes the brightest cusp emission, which radiated 13$\pm$0.5~GW.

SW and planetary rotation both control SKR activity and UV aurorae. While SW-induced magnetospheric compressions trigger global SKR enhancement extending toward low frequencies and lasting for more than a planetary rotation \citep{Desch_JGR_82,Kurth_Nature_05,Lamy_GRL_10,Bunce_JGR_10,Kurth_Icarus_16}, short-lasting SKR intensifications phased with regular SKR bursts are associated with rotationally-modulated nightside injections \citep{Jackman_JGR_09,Mitchell_PSS_09,Lamy_JGR_13,Reed_JGR_18}.

The purpose of Figure \ref{fig2} is to assess the origin of the most active aurorae. Figure \ref{fig2}b-c displays measurements of SKR power and spectral flux density, which quasi-continuously probe the auroral activity at 90~s resolution. Figure \ref{fig2}d-e display SW velocity and dynamic pressure propagated at Saturn. Vertical dashed lines mark the HST observations a-x. We first notice that most of (if not all) long-lasting SKR intensifications with low/high frequency extensions fairly match a SW pressure front within error bars. The five brightest events a, e, l-m and s (boldface dashed lines), with power $\ge$75~GW, peak brightnesses $\ge$150~kR and high latitude emissions, all coincide with such SKR enhancements (see high resolution dynamic spectra in supplementary Figure S3) and can therefore be identified as SW-driven auroral storms. Precisely, the UV observations were respectively acquired $\sim$23, 19, 28-30 and 20~h after the rise of SKR activity which respectively lasted for $\ge$52, 22, 47 and 32~h with multiple bursts, so that the HST images diagnosed a late stage of 4 different storms. 

In contrast, the 2 consecutive images j-k, radiating $\sim70$~GW with peak brightnesses $\ge80$~kR reveal a main emission confined at usual latitudes with a midnight active region which rotated toward dawn. These were acquired during quiet SW conditions at northern SKR phases of 301 and 354$^\circ$, just before a modest northern SKR burst (see supplementary Figure S3) roughly consistent with the arrival of the active region to dawn. The auroral episodes j-k thus suggest a rotationally-driven nightside injection. 



\subsection{Average aurora at solstice}
\label{average}

\begin{figure*}
\centering
\noindent\includegraphics[width=35pc,angle=-0]{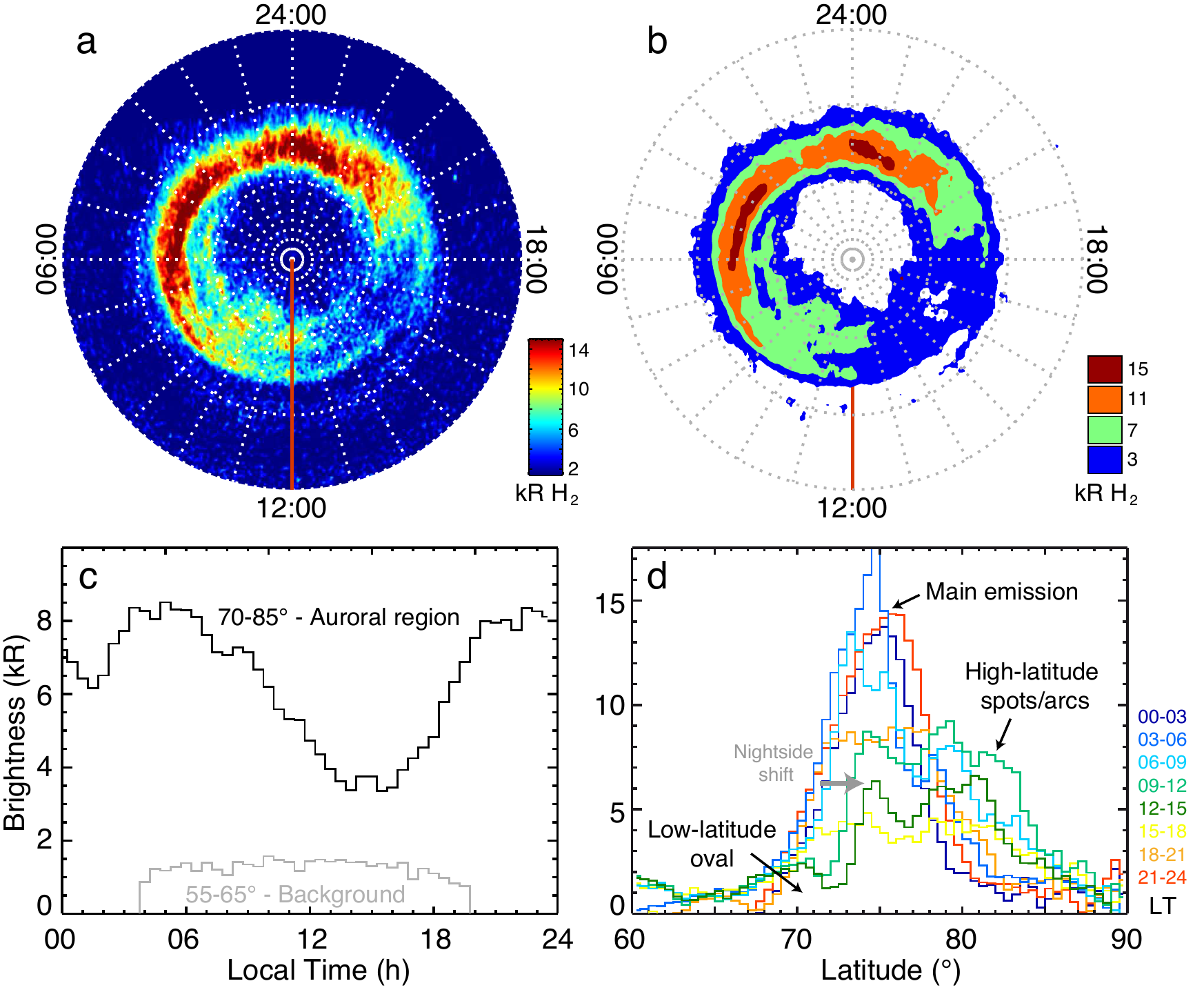}
\caption{(a) Average of the 24 polar projections of Figure \ref{fig1} smoothed with a 3 pix-wide running box. (b) Same as (a) but smoothed with a 17 pix-wide box instead to increase the signal-to-noise, with iso-contours at 3, 7, 11 and 15~kR. Both panels display the average locus and brightness of the circumpolar main oval within $70-85^\circ$ latitude, of high latitude emissions near noon and of a dayside secondary oval near $70^\circ$. (c) Average intensity profiles within $70-85^\circ$ latitude (black line, aurora) and $55-65^\circ$ (gray line, background) versus LT. The auroral intensity peaks at $05$:$00$ and $22$:$00$~LT. (d) Same as (c) versus latitude for successive 02~h-wide (colored) LT ranges. The main emission is visible at all LT, with distinct high and low latitude emissions mostly visible on the dayside.}
\label{fig3}
\end{figure*}

We now turn to the mean spatial distribution and intensity of Saturn's northern UV aurorae. Figures \ref{fig3}a-b displays average polar projections versus LT. Figures \ref{fig3}c-d display average intensity profiles versus LT and latitude. The average main oval is a circumpolar ring of emissions of a few kR confined within $70-80^\circ$ latitudes. Increasing brightnesses then gradually map to a dusk-to-noon partial ring, with the brightest emissions ($15$~kR iso-contour in panel b) between pre-midnight and dawn. 

The main oval expands poleward beyond $\sim80^\circ$ between $04$:$00$ and $21$:$00$~LT, encompassing polar arcs and spots produced by dawnside auroral storms, noon cusp and duskside bifurcations. The low average brightness of these high latitude emissions illustrates their transient nature. Figures \ref{fig3}a-b,d additionally reveal a nightside shift of the main oval (gray arrow in Figure \ref{fig3}d) : while the low-latitude boundary extends down to 70-72$^\circ$ on the nightside, it reaches 73-75$^\circ$ instead between $09$:$00$ and $15$:$00$~LT, with highest latitudes at $14$:$00$~LT.

This shift enables us to identify a distinct equatorward secondary oval within $67-73^\circ$ in Figures \ref{fig3}a-b,d, split from the main oval, with a typical mean brightness of $\sim$2~kR, occasionally exceeding 3~kR (the standard deviation in Figures \ref{fig3}a-b is $\le$1.5~kR). It is pretty remarkable that this secondary oval is detected quasi-continuously from noon to dawn and dusk with roughly homogeneous intensities, which suggest that it primarily consists of steady weak emissions. These characteristics match the 1.7~kR secondary oval previously detected at $-67^\circ$ nightside latitude. When considering the 2$^\circ$ northern latitudinal shift due to the northern magnetic field offset, the northern dayside oval appears at slightly higher latitudes than the nightside southern one, in agreement with the nightside shift of the main oval.

Figure \ref{fig3}c quantifies the LT dependence of northern aurorae discussed above by plotting the average brightness profile over $70-85^\circ$ latitude (black line), a region fully visible at all LT, versus LT. It displays two distinct peaks at $05$:$00$ LT and $22$:$00$ LT. When building the same Figure without SW-driven auroral storms, these peaks appear at $06$:$30$ and $20$:$30$~LT instead.


Finally, we similarly investigated the role of planetary rotation, previously found to modulate both the intensity and position of the main oval in both hemispheres \citep{Nichols_GRL_10a,Nichols_GRL_10b}. Supplementary Figure S4 displays the average brightness over $70-80^\circ$ latitude for each image versus northern SKR phase and LT. The dashed line displays a guide meridian indicating an active auroral region rotating at the northern SKR period and reaching $06:00$~LT at a phase of $0^\circ$ (when SKR bursts occur). No active region can be continuously tracked along or close to this guide meridian, in contrast with previous positive results obtained in the southern hemisphere \citep{Nichols_GRL_10a,Lamy_PRE7_11,Lamy_JGR_13}. Supplementary Figures S5a-b display average polar projections similar to Figures \ref{fig3}a-b but versus northern SKR phase. The LT were transposed into phases by again assuming that SKR maxima occur at the pass of a rotating active region through $06:00$~LT. There is no maximum near $0^\circ$, in agreement with Figure S4. The two maxima observed near 90$^\circ$ and 180$^\circ$ instead mainly result from the SW-driven auroral storms. Interestingly, the low-latitude boundary of the main emission ($e.g.$ 7~kR iso-contour in Figure S5b) clearly reaches higher latitudes on the left-hand side (74-75$^\circ$ latitude at 90$^\circ$ phase) than on the right-hand side (72-73$^\circ$ latitude at 180$^\circ$ phase). This latitudinal shift is roughly consistent with that predicted from the tilt of the northern oval when the upward current layer reaches dawn \citep{Nichols_GRL_10b} and therefore suggests a rotational control of the oval's position and not of its intensity, as already observed between 2011 and 2013 \citep{Nichols_Icarus_16}.

\section{Discussion}
\label{discussion}

In the previous section, we described the individual and average properties of the kronian northern aurorae at solstice, some of which are further discussed below.

The peak of northern average auroral brightness at $05$:$00$~LT matches that previously seen with Cassini/UVIS observations \citep{Carbary_JGR_12}, although its displacement toward $06$:$30$ when removing the auroral storms provides a typical uncertainty linked to the statistics of the dataset. This may account for the discrepancy between the southern average auroral brightness peak observed by HST STIS/ACS at $09$:$00$ \citep{Lamy_JGR_09} and by Cassini/UVIS at $06$:$00$~LT \citep{Carbary_JGR_12}. More importantly, the HST observations of 2017 additionally reveal a second peak of comparable amplitude at $20:30-22:00$~LT, previously unreported, and strikingly reminiscent of Earth's aurora. This LT sector is also the one where the equatorward latitude of the main oval minimizes ($\ge3$~kR iso-contours in Figure \ref{fig3}b). This secondary peak may thus arise either from a better viewing of the nightside sector or from more frequent nightside injections than previously observed, possibly favored under solstice conditions.

In contrast with a clear LT dependence of the auroral intensity, the rotational dynamics do not seem to play a significant role, if we except the moderate brightening phased with the northern SKR seen in the two successive images j-k and clues of a rotational control of the oval's position. Instead, the most obvious variability is that induced by large scale auroral storms driven by the SW, for which SKR quasi-continuous observations provide total durations estimated between 22 and 48~h at least. This largely exceeds the $11-21$~h range previously inferred by \citep{Meredith_JGR_13}, which explained a typical duration of $\sim1.5$ planetary rotations for the time needed by hot plasma injected from the nightside with a 60\% subcorotational motion to complete one rotation. Accounting for the observed durations would imply very low sub-corotational rates (from 20 to 50\%). Instead, we propose that storms generally do not result from a single magnetospheric compression but from a series of them consistent with the multiple SKR bursts observed along each long event. The peculiar spiral shape observed in image a with extremely high latitudes surrounding the northern magnetic pole finally questions the interpretation of the main emission as a tracer of the open-closed field line boundary located 1-2$^\circ$ poleward, in which case the polar cap would correspond to a very small region around the pole. Analyzing Cassini {\it in situ} measurements obtained a few hours before image a, when the spacecraft sampled the dawnside region poleward of the main oval, is required to adress this question. The detailed study of another example of SW-driven UV auroral storm temporally resolved by Cassini/UVIS with a concurrent SKR long-lasting enhancement is the subject of a companion paper \citep{Palmaerts_GRL_18}.


The apparent systematic $\sim3^\circ$ nightside shift in latitude of the average northern auroral oval, previously unreported to our knowledge, is consistent with the modeled influence of the SW flow on the open-closed field line boundary \citep{Belenkaya_AG_08}.

Cusp emissions and bifurcations were observed very frequently, in $\sim50\%$ of the images. This unusually high occurrence rate is likely related to the magnetosphere/SW configuration reached at solstice. This scenario implies enhanced SW-driven mass loading of the magnetosphere, and therefore supports more frequent nightside plasmoid releases/injections. The observation of an unusually bright cusp emission in image s, extended by a duskside bifurcation connecting it to a spiral-shaped main oval and observed during an auroral storm, suggests that it may have been induced by a SW-driven magnetospheric compression as observed at Earth \citep{Farrugia_JGR_95} and similarly proposed at Uranus \citep{Lamy_JGR_17}.

The novel identification of a dayside low latitude emission with $2-3$~kR brightnesses on average (which compare to that previously identified with comparable brightness and latitude on the nightside) imply a steady mechanism able to operate at all longitudes. In addition, it is worth noting that the dayside portion of the oval is not seen in all individual images and in half of the cases during auroral storms. This suggests an additional transient activity possibly linked to dayside compression of magnetic field lines. The previously proposed origin of such emission related to a suprathermal population of electrons in the middle magnetosphere \citet{Grodent_JGR_10} is called into question by both this transient activity and by {\it in situ} magnetic measurements consistent with small-scale field-aligned currents. These features are alternatively consistent with either ancient long-lived structures of the main emission associated with strong field-aligned currents which moved toward lower latitudes and/or auroral precipitations associated with a secondary current system previously observed within $68.5-72^\circ$ latitudes \citep{Hunt_JGR_15b}. The detailed analysis of Cassini plasma measurements during this peculiar event is necessary to address this question.

\section{Conclusion}

In this study, we analyzed 24 HST/STIS images of Saturn's northern aurorae acquired throughout 2017 during northern summer solstice, concurrently with Cassini/RPWS observations of SKR and numerically propagated SW parameters. The observed northern aurorae display highly variable auroral components, with a total power ranging from 7 to 124$\pm11$~GW. The prominent component is the main oval observed poleward of 72$^\circ$ and shifted by $\sim3^\circ$ toward the nightside which bears clear signatures of the SW (4 auroral storms coincident with SKR long-lasting enhancements) and planetary rotation (1 auroral brightening coincident with a regular SKR burst). Recurrent cusp emissions and bifurcations are unexpectedly frequent, in 50\% of the images, with an unusually bright cusp emission observed during an auroral storm which radiated $13\pm1$~GW, likely triggered by the SW. The identification of a dayside secondary oval at $70^\circ$ latitudes, $2-3$~kR bright on average with some clues of temporal variability brings new constraints to its possible origins. On average, the northern solstice aurorae display a strong LT dependence with two maxima at dawn and pre-midnight, the latter being attributed to regular nightside injections, with clues of a rotational control of the oval's average position, but not of its intensity. These results provide a reference frame to analyze Cassini {\it in situ} measurements, whether simultaneous (the Cassini footprints intercepted an auroral component in half of the images) or not.

%

\acknowledgments
The UV observations were obtained from the ESA/NASA Hubble Space Telescope (GO program \#14811) : the original data can be retrieved from the MAST archive and the processed data from the APIS service hosted by the Paris Astronomical Data Centre at \url{http://apis.obspm.fr}. The Cassini/RPWS and MAG original data are accessible through the PDS archive at \url{https://pds.nasa.gov/}. The HFR processed data are available through the LESIA/Kronos database at \url{http://www.lesia.obspm.fr/kronos}. LL thanks Linda Spilker for her support to the original HST proposal, Fannie Serrano and Pauline Richard, who investigated short-term dynamics of Saturn's aurorae during their internship at LESIA, and Gabby Provan for useful discussions on the inexhaustible topic of kronian magnetospheric periodicities. The French co-authors acknowledge support from CNES and CNRS/INSU programs of Planetology (PNP) and Heliophysics (PNST). SVB was supported by an STFC Ernest Rutherford Fellowship ST/M005534/1. The research at the University of Iowa was supported byÊNASAÊthroughÊContractÊ1415150 with the Jet Propulsion Laboratory.

%
%
%
%
%
%
%
%
%





\end{document}